\documentstyle[twocolumn,seceq,epsbox]{jpsj}

\def\runtitle
{
  Thermodynamic Properties 
  of Three-Dimensional Orthogonal Dimer Model for SrCu$_2$(BO$_3$)$_2$
}
\def\runauthor{
  Shin {\sc Miyahara}
  and Kazuo {\sc Ueda}
}

\title
{
  Thermodynamic Properties 
  of Three-Dimensional Orthogonal Dimer Model for SrCu$_2$(BO$_3$)$_2$\\
}

\author
{
  Shin {\sc Miyahara}
  and Kazuo {\sc Ueda}
}

\inst
{
Institute for Solid State Physics, University of Tokyo,
  Roppongi, Tokyo 106-8666, Japan
}

\recdate
{
  \today
}

\abst {
  Effects of the inter-layer couplings
  for the orthogonal dimer system
  SrCu$_2$(BO$_3$)$_2$ are discussed.
  The spin-gap $\Delta$ of the three-dimensional model
  is independent of the inter-layer couplings when they are small.
  Therefore at low temperatures ($T < \Delta$) thermodynamic
  properties are described well by the two-dimensional model.
  On the other hand at high temperatures the mean-field type
  scaling ansatz is useful to discuss the magnetic susceptibility
  for week inter-layer couplings.
  From fit of the magnetic susceptibility,
  the estimated coupling constants are $J = 85$ K for the
  nearest-neighbor couplings, $J^{'} = 54$ K for the
  next-nearest-neighbor couplings, and $J^{''} = 8$ K
  for the inter-layer couplings.
  These parameters are consistent with the temperature dependence of
  the specific heat at low temperatures.
  }

\kword
{
  SrCu$_2$(BO$_3$)$_2$, uniform susceptibility,spin-gap, specific heat,
  inter-layer coupling, orthogonal dimer state, Shastry-Sutherland model
}

\begin{document}
\makeatletter
\def\lessim{\mathrel{\mathpalette\gl@align<}}
\def\gtrsim{\mathrel{\mathpalette\gl@align>}}
\def\gl@align#1#2{\lower.6ex\vbox{\baselineskip\z@skip\lineskip\z@
    \ialign{$\m@th#1\hfil##\hfil$\crcr#2\crcr\sim\crcr}}}
\makeatother

\maketitle

\section{Introduction}

\begin{figure}[hbt]
  \begin{center}
    \psbox[width=6.5cm]{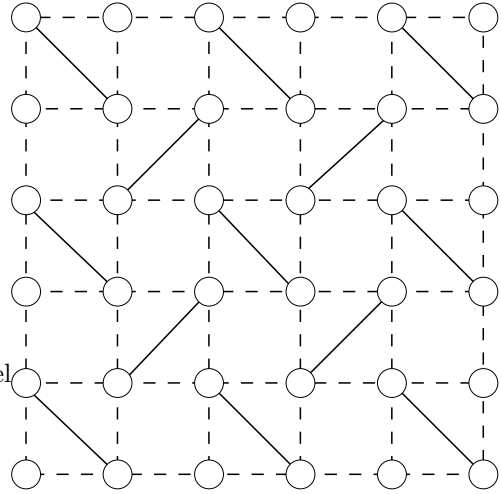}
  \end{center}
  \caption{The model discussed by Shastry and Sutherland,
    which has the dimer singlet ground state.}
  \label{fig:S-Smodel}
\end{figure}

The new spin-gap system SrCu$_2$(BO$_3$)$_2$,
which was found by Kageyama {\it et al}.~\cite{kageyama},
is a beautiful realization of
the two-dimensional Shastry-Sutherland model
(Fig.~\ref{fig:S-Smodel}),
which was studied by them almost
twenty years ago~\cite{shastry}.
The original model seems to be very artificial.
In fact, it was constructed in such a way to realize
an exact ground state on a two-dimensional model.
The system for SrCu$_2$(BO$_3$)$_2$ is shown
in Fig.~\ref{fig:lattice_2d}(a), which is described as
the orthogonal dimer model.~\cite{kageyama,miyahara1}
It is topologically equivalent
to the two-dimensional Shastry-Sutherland model,
where the nearest-neighbor bond in the former model corresponds to
the next-nearest-neighbor bond of the latter and vice versa.
The orthogonal dimer model
shows a quantum phase transition from
the gapful phase to the gapless phase 
as the next-nearest-neighbor interaction is increased~\cite{koga}.
In the gapful phase the ground state is
the exact dimer ground state. 

SrCu$_2$(BO$_3$)$_2$ consists of the layers of CuBO$_3$ and
the layers of Sr. In the compounds Cu$^{2+}$ ions occupy
crystallographically equivalent sites and described by
a spin-$1/2$ Heisenberg model.
As the first approximation the compound may be treated
as a two-dimensional system and the novel features of the
SrCu$_2$(BO$_3$)$_2$, for example
the magnetization plateaus,
are explained by this model.
In fact the $1/3$-plateau predicted
by theory~\cite{momoi,miyahara2,hukumoto}
was observed in the recent experiment
in high magnetic fields~\cite{onizuka}. 
The coupling constants were estimated from the spin-gap
and the Curie-Weiss constant.
According to the estimation, SrCu$_2$(BO$_3$)$_2$ is
regarded as a material which is near
the quantum phase transition point~\cite{miyahara1}.
Originally, the spin-gap observed in the powder sample $30$ K
was used.
While, more recent experiments using the single crystal samples
show that the gap is close to $35$ K~\cite{kageyama2,nojiri,kageyama4}.

In this paper we will study the thermodynamic properties of
SrCu$_2$(BO$_3$)$_2$ in more detail and reestimate the coupling
constants for this material taking also
the inter layer couplings into consideration.
First, one can show that
the ground state and the first excited state
are independent of the interlayer couplings.
For that reason the properties of SrCu$_2$(BO$_3$)$_2$
at low temperatures can be described well by the two-dimensional model.
Therefore we discuss the temperature dependence
of the magnetic susceptibility based
on the two-dimensional model at low temperatures
and determine the coupling constants
in the CuBO$_3$ plane.
Then we show that these parameters are consistent with the specific heat.
From the susceptibility at high temperatures
the inter-layer couplings may be estimated.

\section{Ground state and first excited state}

\subsection{Two-dimensional orthogonal dimer model} 

\begin{figure}[hbt]
  \begin{center}
    \psbox[width=7.5cm]{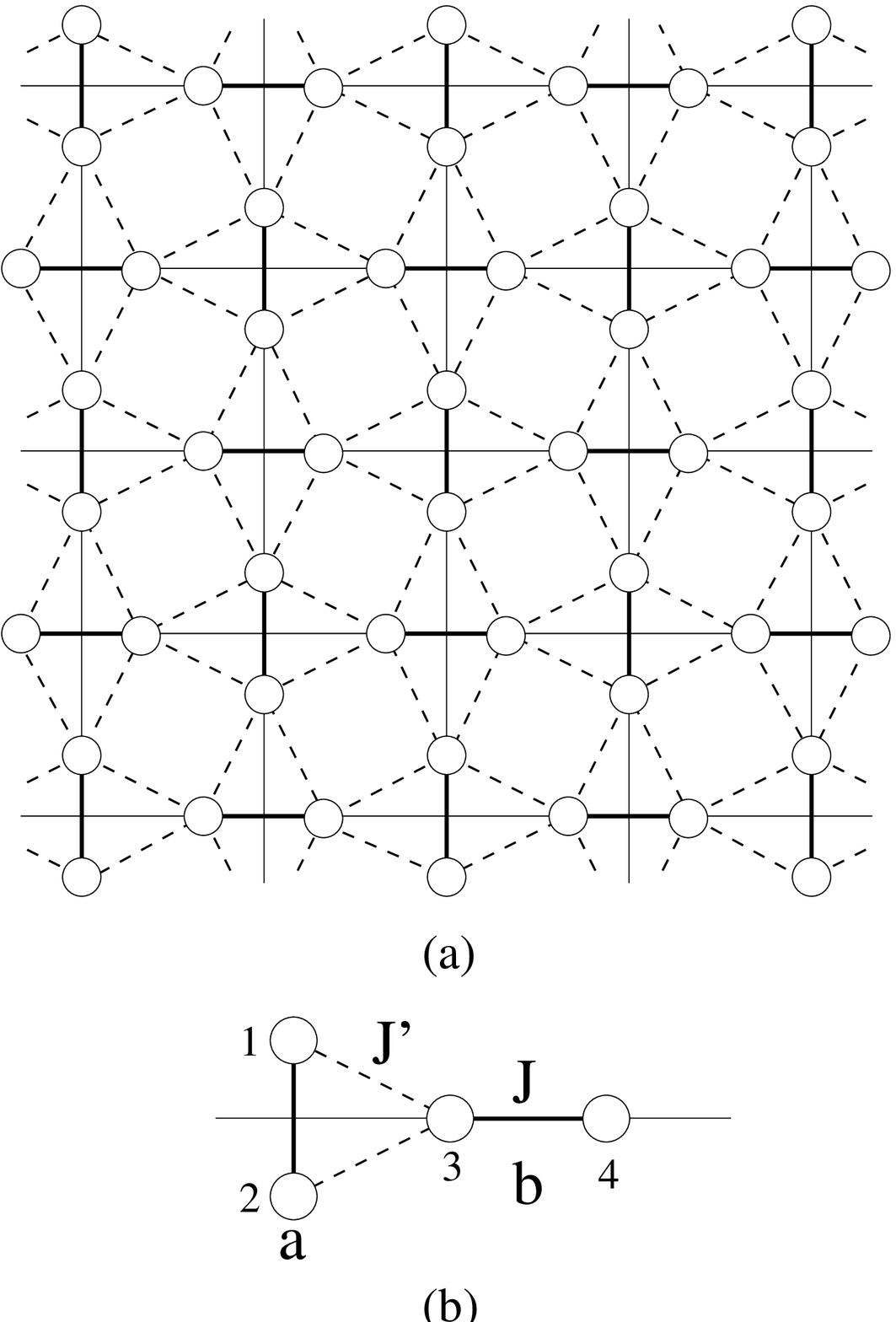}
  \end{center}
  \caption{(a)The model for SrCu$_2$(BO$_3$)$_2$ : Two-dimensional
    orthogonal dimer model.
    (b)A configuration of two dimers which are orthogonal.}
  \label{fig:lattice_2d}
\end{figure}

As shown in our previous analysis~\cite{miyahara1}
the magnetic properties of SrCu$_2$(BO$_3$)$_2$
are described rather well by the two-dimensional Hamiltonian:
\begin{equation}
  {\cal H} =J\sum_{\rm n.n.} {\bf s}_i \cdot {\bf s}_j
  +J^{'}\sum_{\rm n.n.n.} {\bf s}_i \cdot {\bf s}_j\ .
  \label{eq:model}
\end{equation}
The system is shown in Fig.~\ref{fig:lattice_2d} (a).
The model can be considered as a coupled dimer model.
The dimers, where two spins are coupled
with the nearest-neighbor coupling $J$,
are connected by the next-nearest-neighbor bond $J'$.
An elementary unit for the interaction between a pair of the
dimer bonds is shown in Fig.~\ref{fig:lattice_2d} (b).
It is convenient to use the dimer bases defined for 
each nearest-neighbor bond:
\begin{eqnarray}
   | s\rangle & =&\frac{1}{\sqrt{2}}(| \uparrow \downarrow \rangle -
        |  \downarrow \uparrow \rangle ), \\
   | t_1 \rangle & =& | \uparrow \uparrow \rangle, \\
   | t_0 \rangle & =&\frac{1}{\sqrt{2}}(| \uparrow \downarrow \rangle +
        |  \downarrow \uparrow \rangle ), \\
   | t_{-1} \rangle & = &| \downarrow \downarrow \rangle.
\end{eqnarray}\\

The direct product of the singlets on dimers defined by 
\begin{equation}
   |\Psi\rangle = \prod_a | s\rangle_a
\end{equation}
is an exact eigenstate of the Hamiltonian
(\ref{eq:model})~\cite{shastry,miyahara1}.
Here the index $a$ denotes each dimer bonds and
runs over all dimer bonds.
For coupling constants $J'/J < 0.69$ this eigenstate
is the ground state~\cite{miyahara1,weihong,muller}.
This model shows a quantum phase transition at $J'/J = 0.69$
from the dimer singlet state
to the N\'{e}el ordered state which is gapless~\cite{koga}.
Note that in the limit of $J^{'}/J \rightarrow \infty$,
the present model reduces to the square lattice Heisenberg model,
whose coupling constant is $J^{'}$.

\begin{figure}[hbt]
  \begin{center}
    \psbox[width=7.5cm]{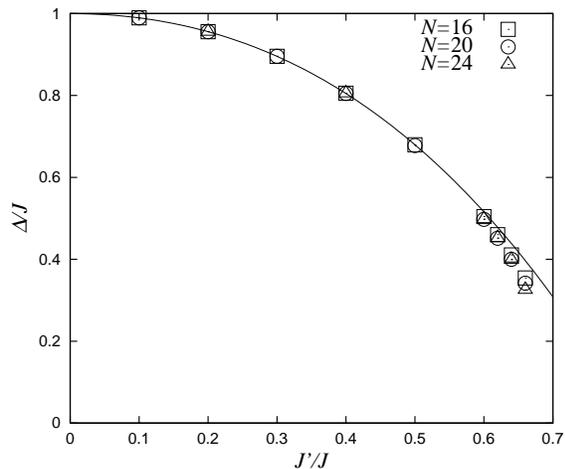}
  \end{center}
  \caption{
    Spin-gap for finite lattices: $N_s = 16$, $20$, and $24$
    from the dimer singlet ground state.
    The solid line is the perturbation result up
    to the fourth order.}
  \label{fig:spingap}
\end{figure}

The singlet dimer ground state
has a spin-gap, which can be estimated by
the perturbation theory.
The spin-gap up to the fourth-order is given by
\begin{equation}
  \frac{\Delta}{J} = 1 - (\frac{J^{'}}{J})^2 -
  \frac{1}{2}(\frac{J^{'}}{J})^3 - \frac{1}{8}(\frac{J^{'}}{J})^4,
  \label{eq:gap}
\end{equation}
and the result up to the fifteenth-order is given
in Ref.~\citen{weihong}.
The spin-gap for finite systems is shown in Fig. \ref{fig:spingap},
where the number of the spins is $16$, $20$, and $24$ with
periodic boundary conditions.
The finite size effects for $J'/J \lessim 0.66$ are small.
The results of eq.~(\ref{eq:gap}) and the numerical results for
the finite size systems agree well for $J'/J < 0.5$.\\

The perturbation theory predicts a novel character for
the triplet excitation.
The triplet excitation is completely localized
up to the fifth-order, which leads to crystallization
of the triplet excitations at certain magnetizations.
At magnetizations where the crystallization occurs,
the magnetization plateaus appear~\cite{miyahara1,miyahara2,momoi,hukumoto}.
This feature of the triplet excitations can be understood
from the matrix elements for one triplet excitation.
\begin{equation}
  J^{'} ( {\bf s}_1 + {\bf s}_2) \cdot {\bf s}_3
  | t_1\rangle_a | s\rangle_b =
  \frac{J^{'}}{2}| t_1\rangle_a | t_0\rangle_b
  - \frac{J^{'}}{2}| t_0\rangle_a | t_1\rangle_b~,
  \label{eq:triplet_matrix1}
\end{equation}
and
\begin{equation}
  J^{'} ( {\bf s}_1 + {\bf s}_2) \cdot {\bf s}_3
  | s\rangle_a | t_m\rangle_b  = 0
  \hspace{0.5cm}(m = 0, \pm1)~,
  \label{eq:triplet_matrix2} 
\end{equation}
where the site indices are specified in Fig.~\ref{fig:lattice_2d}(b).
Equation ~(\ref{eq:triplet_matrix1}) means that when a triplet moves
to one of neighboring bonds it leaves another triplet behind.
From the symmetry reason, the parity with the reflection,
the matrix element shown in eq.~(\ref{eq:triplet_matrix2}) vanishes.
Equations.~(\ref{eq:triplet_matrix1}) and (\ref{eq:triplet_matrix2})
make a hopping of a triplet excitation rather difficult.
It becomes possible through a closed path of dimer bonds
and thus the hopping processes  start from the sixth-order
in the perturbation.
Recently this almost localized nature of the dispersion is
directly observed in the
neutron scattering experiment~\cite{kageyama4}. 



\subsection{Effects of the inter-layer couplings}

\begin{figure}[hst]
  \begin{center}
    \psbox[width=7.5cm]{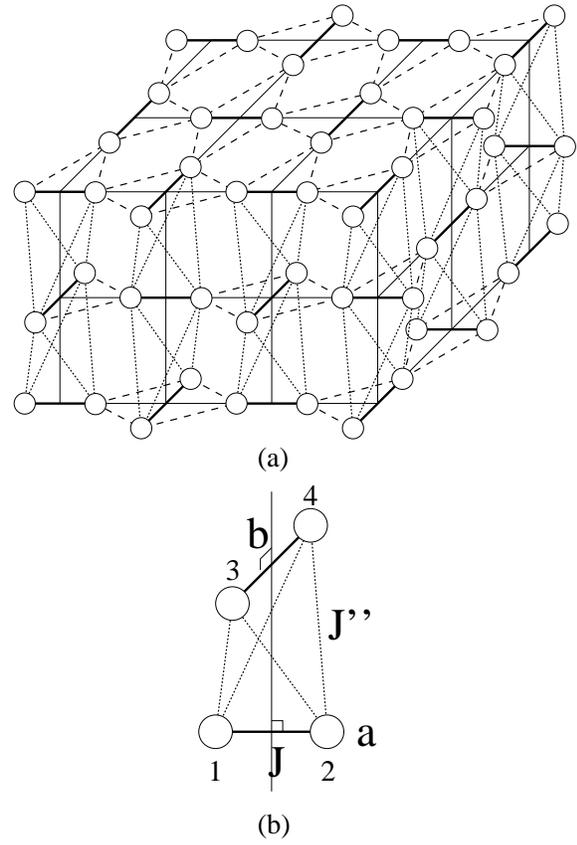}
  \end{center}
  \caption{
    (a) The model for SrCu$_2$(BO$_3$)$_2$ : Three-dimensional
    orthogonal dimer model.
    (b) The configurations of two dimers along the $c$-axis
    are also orthogonal in a different way
    from Fig.~\ref{fig:lattice_2d}(b).}
  \label{fig:lattice_3d}
\end{figure}

As we noted before,
SrCu$_2$(BO$_3$)$_2$ consists of CuBO$_3$ layers
and Sr layers. The CuBO$_3$ layers stack alternately
as is shown in Fig.~\ref{fig:lattice_3d}(a).
Along the $c$-axis dimers are coupled with inter-layer coupling $J^{''}$
as shown in Fig.~\ref{fig:lattice_3d}(b).
It is obvious that the matrix element for the singlet dimers
vanishes:
\begin{equation}
  J^{''} ( {\bf s}_1 + {\bf s}_2) \cdot
  ( {\bf s}_3 + {\bf s}_4) | s \rangle_a | s \rangle_b  =0\ .
\end{equation}
Therefore the three-dimensional model for SrCu$_2$(BO$_3$)$_2$
has the exact orthogonal dimer ground state for
small $J^{'}$ and $J^{''}$.~\cite{ueda}\\

Next, let us consider the case where a triplet is excited.
The matrix element from this state also vanishes:
\begin{equation}
  J^{''} ( {\bf s}_1 + {\bf s}_2) \cdot
  ( {\bf s}_3 + {\bf s}_4) | t_m \rangle_a | s \rangle_b  =0\
  \hspace{0.5cm}(m = 0, \pm1) .
\end{equation}
It is obvious that the triplet excitation cannot move along
the $c$-axis at low temperatures.
If the neighboring planes are filled with the dimer singlet states,
the triplet excitation on a plane is completely confined in that plane.
Thus the magnitude of the spin-gap for the three-dimensional model
is the same as the two-dimensional one.
The dispersion for the triplet excitation
is not modified, either.

The inter-layer coupling $J^{''}$ does not affect the spin-gap.
It means that the properties of SrCu$_2$(BO$_3$)$_2$
can be described by using the two-dimensional orthogonal dimer model
at low temperatures: $T < \Delta$, even if the inter-layer
couplings exist.
Notice that the theoretical predictions discussed
so far~\cite{miyahara1,miyahara2,momoi,hukumoto} are hardly changed
at low temperatures
by the inclusion of finite inter-layer couplings, which are smaller
than $J'$.

  


\section{Thermodynamic properties of SrCu$_2$(BO$_3$)$_2$}

\subsection{Magnetic susceptibility}

Here we discuss the temperature dependence of the
magnetic susceptibility of SrCu$_2$(BO$_3$)$_2$ 
and determine an optimal set of the parameters
$J$, $J^{'}$, and $J^{''}$.

\begin{figure}[hbt]
  \begin{center}
    \psbox[width=7.5cm]{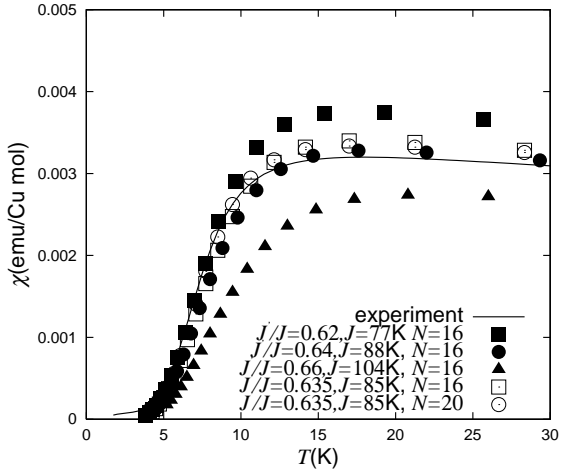}
  \end{center}
  \caption{
    The temperature dependence of the magnetic susceptibility
    on the two-dimensional model. $J^{'}/J = 0.62$,
    $J^{'}/J = 0.64$, $J^{'}/J = 0.635$, and $J^{'}/J = 0.66$. 
    }
  \label{fig:susc}
\end{figure}

Let us start from low temperatures: $T < \Delta$. 
The magnetic susceptibility at $T < \Delta$
may be described well by the susceptibility
of the two-dimensional model $\chi_{2d}$.
Therefore we calculate $\chi_{2d}$ with the transfer matrix method
and compare the results with experimental ones at $T < 30$ K.
The results are shown in Fig.~\ref{fig:susc}.
The results for $J^{'}/J = 0.62, 0.64, 0.66$ are shown by
the filled symbols.
The spin-gap $\Delta \sim 35$ K 
is obtained by various experiments~\cite{kageyama2,kageyama4,nojiri},
and $J$ is determined so that the spin-gap obtained
from the exact diagonalization for $N_s =24$ is $35$ K:
$J = 72$ K for $J^{'}/J = 0.62$,  $J = 88$ K for $J^{'}/J = 0.64$,
and $J = 104$ K for $J^{'}/J = 0.66$.  
Here we note that if we determine the parameters with which both
the Curie-Weiss constant and the spin-gap is satisfied
by the two-dimensional model,
we get $J^{'}/J = 0.66$ and $J = 104$ K.
The system size $N_s = 16$ is used for the calculations
at finite temperatures.
In the present system the finite size effects is not so important
because the triplet excitation is almost localize.
In Fig.~\ref{fig:susc}, to show finite size effects,
the results for $N_s = 16$ and $20$
are shown with open symbols for $J^{'}/J = 0.635$,
which is considered to be an optimal choice.
The finite size effects are observed but small.
Considering the finite size effects,
we may conclude that the parameters
$J^{'}/J = 0.635 \pm 0.01$ explain the experiment well.
In the following, we use the parameters $J^{'}/J = 0.635$ and $J = 85$K
as the optimal set for SrCu$_2$(BO$_3$)$_2$.

\begin{figure}[hbt]
  \begin{center}
    \psbox[width=7.5cm]{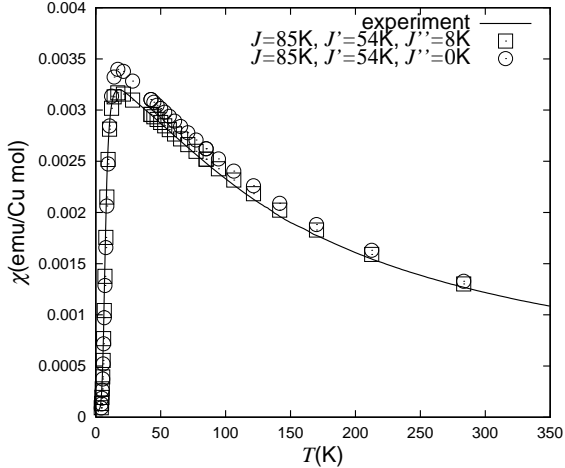}
  \end{center}
  \caption{
    Fit of the uniform magnetic susceptibility for
    SrCu$_2$(BO$_3$)$_2$.
    The parameters used are $J = 85$ K, $J^{'} = 54$ K, and $J^{''} = 8$ K.
    }
  \label{fig:susc_mean}
\end{figure}

Now we consider the magnetic susceptibility
at temperatures $T > \Delta$ .
As is shown in Fig.~\ref{fig:susc_mean},
the calculated susceptibility for $J^{'}/J = 0.635$ and $J = 85$K,
which agrees with experiments well at $T < \Delta$,
does not fit to the experiment
in the temperature range: $20 < T < 350$.
The system size effects at the temperatures $T > 50$ K is
so small that the results for $N = 20$
are considered as the bulk limit
for such temperature range.
This fact shows that
the effects of the inter-layer couplings $J^{''}$ on the susceptibility
cannot be ignored in such a temperature range.
To estimate the magnitude of $J^{''}$, we follow the mean-field
type scaling ansatz used in Ref.~\citen{johnstone}:
\begin{equation}
  \chi(J^{'}/J, J^{''}/J) = \frac{\chi_{2d}(J^{'}/J)}
  {1 + 4 J^{''} \chi_{2d}(J^{'}/J)}.
  \label{eq:meanfield}
\end{equation}
The coefficient $4 J^{''}$ in the denominator reproduces
the correct high-temperature
Curie-Weiss constant of the three-dimensional model.
At low temperatures this ansatz gives the same spin-gap
as that given by the two-dimensional model, which is
reasonable because the spin-gap is not modified by
a small $J^{''}$.

\begin{figure}[hbt]
  \begin{center}
    \psbox[width=7.5cm]{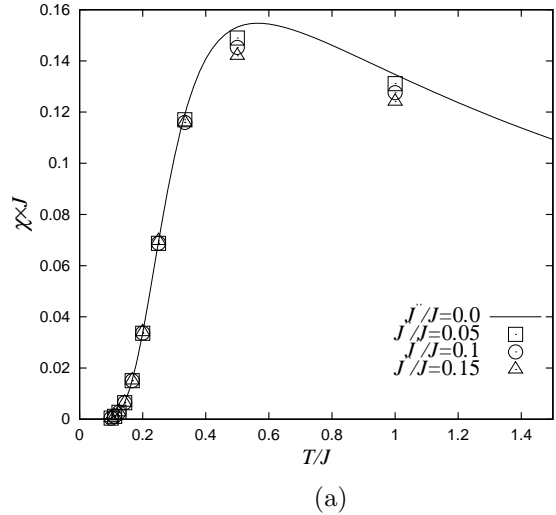}

    (a)
  \end{center}
  
  \begin{center}
    \psbox[width=7.5cm]{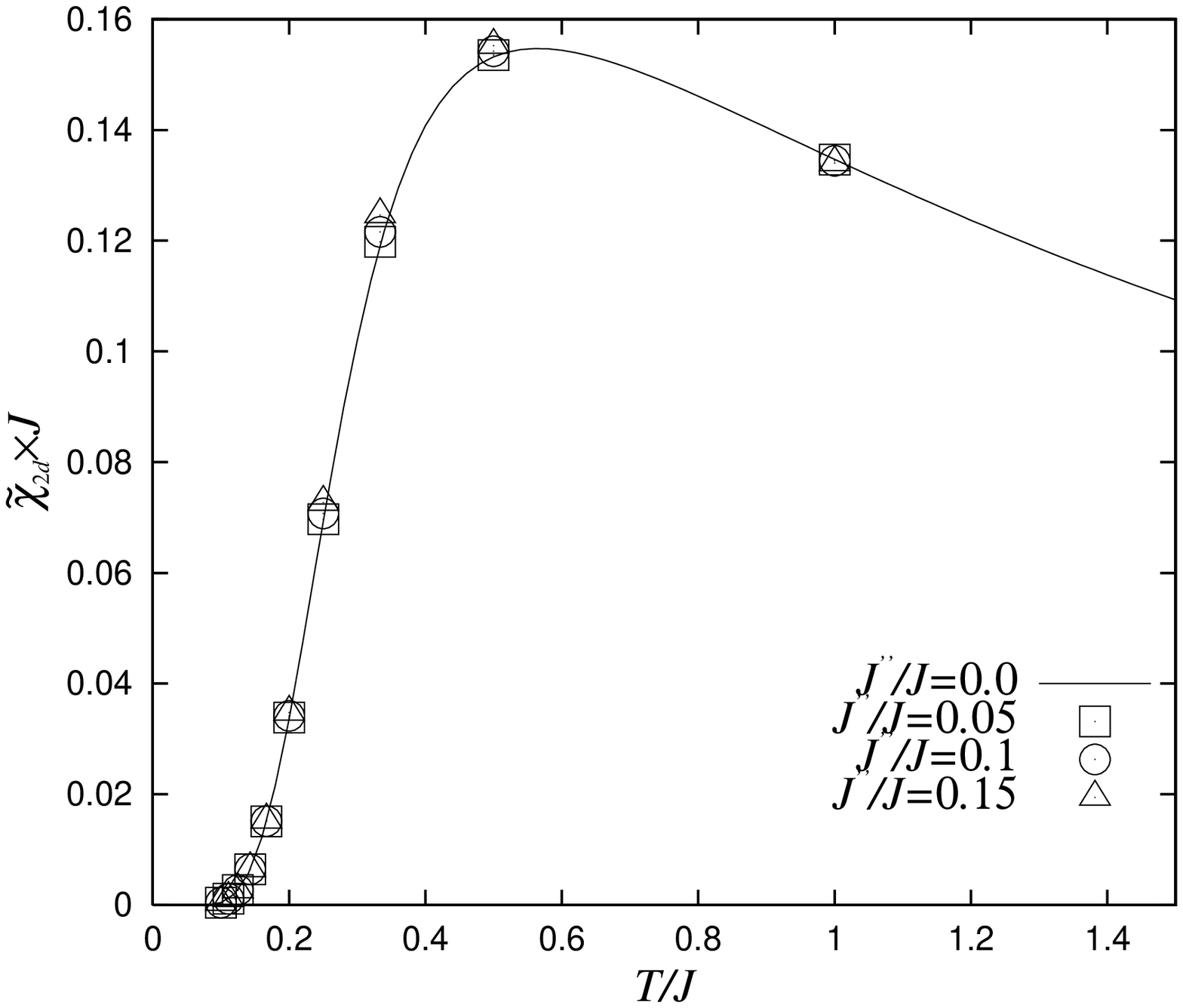}

    (b)
  \end{center}
  \caption{
    (a) The temperature dependence of the magnetic susceptibility
    of the bilayer model. $J^{'}/J = 0.4$ and
    $J^{''}/J = 0.05$, $J^{'}/J = 0.1$, and $J^{'}/J = 0.15$.
    The solid line is the results of the two-dimensional model.
    (b) Scaling plot of the mean-field type ansatz.
    }
  \label{fig:susc_3d}
\end{figure}

To check the quality of eq.~(\ref{eq:meanfield})
we have calculated the temperature dependence of
the susceptibility using a bilayer model,
where there are $8$ spins in each layer.
We assume the periodic boundary conditions.
The used parameters are $J^{'}/J = 0.4$ for the in-plane
couplings and $J^{''}/J = 0.05$, $0.1$, $0.15$
for the inter-layer couplings.
The results are shown in Fig.~\ref{fig:susc_3d}(a).
For this system, the spin-gap is $J$ because
of the periodic boundary condition.
We select small  $J^{''}/J$ so that inter-layer couplings
do not change the magnitude of the spin-gap and the ground state.
The ground state and the first excited state for these systems
are checked by the exact diagonalization calculations.
We scale the obtained susceptibility
onto the effective two-dimensional one $\tilde{\chi}_{2d}$
through the inverse relation:
\begin{equation}
  \tilde{\chi}_{2d}(J^{'}/J, J^{''}/J)) = \frac{\chi_{3d}(J^{'}/J, J^{''}/J)}
  {1 - 4 J^{''} \chi_{3d}(J^{'}/J, J^{''}/J)}.
  \label{eq:meanrev}
\end{equation}
The scaling plots are shown in Fig~\ref{fig:susc_3d}(b).
The results form approximately a single line, supporting
the mean-field type ansatz.

Next, using eq.~(\ref{eq:meanfield}),
we estimate the magnitude of the inter-layer coupling $J^{''}$.
Here we use  $\chi_{2d}(J^{'}/J=0.635)$ with $J = 85$ K
because it agrees with the experiments well at low temperatures.
We fit the susceptibility data in the range $100$ K $< T < 350$ K
to eq.~(\ref{eq:meanfield}) and determine $J^{''}$.
Good agreement is obtained for $J^{''}/J = 0.09$.
This fit is shown in Fig.~\ref{fig:susc_mean}:
$J = 85$ K, $J^{'} = 54$ K, and $J^{''} = 8$ K.
The inter-layer coupling $J^{''}$ is about $10\%$ of
the intra-dimer coupling $J$. The inter-layer coupling
is not important at low temperatures but
is important to fix the energy scale of the coupling constants.
One of the origin of the rather big Curie-Weiss constant
$\sim 83$ K may be attributed to 
the existence of the inter-layer couplings. \\



\subsection{Specific heat}

\begin{figure}[hbt]
  \begin{center}
    \psbox[width=7.5cm]{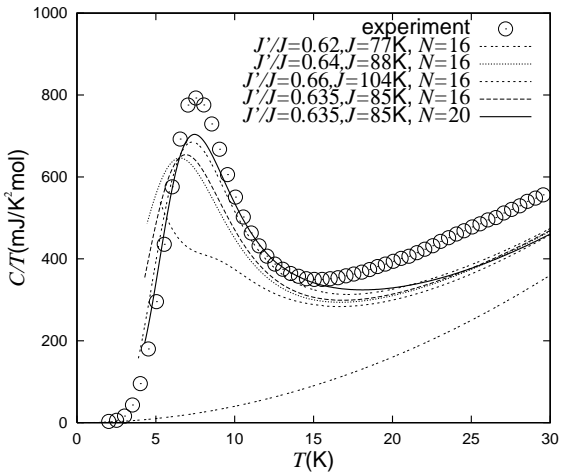}
  \end{center}
  \caption{
    The temperature dependence of the specific heat of
    SrCu$_2$(BO$_3$)$_2$.
    The fit parameters are $J = 85$ K, $J^{'} = 54$ K.
    $0.4 \times T^3$ is used as the contribution from phonon
    degrees of freedom.
    }
  \label{fig:specific}
\end{figure}

Next we discuss the specific heat.
The temperature dependence of the specific heat
of SrCu$_2$(BO$_3$)$_2$ at low temperatures $T < \Delta$
may be also explained well by using the two-dimensional model.
The specific heat is calculated by the transfer matrix methods again.
We compare the theoretical results for $J^{'}/J = 0.635$ and $J = 85$ K,
which explained the temperature
dependence of the susceptibility well, with
the experiments~\cite{kageyama3}.
Here $\beta \times T^3$ is used as
the specific heat for the phonon term.
$\beta = 0.4$ mJ/K$^4$ is used.
The results are shown in Fig.~\ref{fig:specific}.
We see that the specific heat for $J^{'}/J = 0.635$
and $J = 85$ K agrees well with the experiments
and the difference shown around the peak may be explained
by the finite size effects.
In  Fig.~\ref{fig:specific} the specific heat for
$J^{'}/J = 0.62$, $J^{'}/J = 0.64$
and $J^{'}/J = 0.66$ are also shown for comparison.\\

At temperatures $T \gtrsim 15$ K
good quality of the fits is not obtained and
the differences cannot be attributed only
to the system size effects.
Possible reasons of this difference may be the followings.
First, in this temperature range the simple $\beta \times T^3$ expression 
is not sufficient for the lattice contribution.
Second, the effects of the spin-phonon
coupling may be important.
In the orthogonal dimer model, the small kinetic energies
of the excited triplets originate from the geometrical constraint.
When the orthogonality is broken by some distortion of dimer bonds,
a finite matrix element for a hopping of a triplet arises.
Therefore a triplet excitation is expected to be
a strongly coupled with phonons.
Effects of the spin-phonon couplings will play an important role
to understand the behavior of the specific heat,
which is an interesting future problem.


\section{Conclusion}

In the present study we have discussed
thermodynamic properties of SrCu$_2$(BO$_3$)$_2$
with the three-dimensional orthogonal dimer model.
At low temperatures this three-dimensional model
is effectively equivalent to the two-dimensional model
because the spin-gap is not affected by
the inter-layer coupling $J^{''}$.
Therefore SrCu$_2$(BO$_3$)$_2$ can be described well by
using the two-dimensional model as the first approximation,
not because of the weakness of the inter-layer coupling $J^{''}$
but by the geometrical reason.

Even if many triplet excitations exist in one plane,
the triplets move only in the plane 
in spite of the sizable inter-layer couplings
if the dimer bonds on the neighboring planes are occupied
by the singlets.
Therefore the two-dimensional spin dynamics are expected
in SrCu$_2$(BO$_3$)$_2$ even if there are sizable inter-layer
couplings. 
In fact unusual diffusive spin dynamics of two-dimensional character
is observed by the NMR measurements~\cite{takigawa}.

Concerning the coupling constant: $J$, $J^{'}$, and $J^{''}$,
the best fit is given by $J = 85$ K, $J^{'} = 54$ K,
and $J^{''} = 8$ K.
In the real system the horizontal and vertical bonds
in each plane are shifted slightly,
which means that the one type of dimers,
for example horizontal dimers,
are on a plane and the other dimers, vertical dimers, are on
the other plane. Therefore the inter-layer couplings along
the $c$-axis consist of the two coupling constants.
Here we treat $J^{''}$ by the mean-field type
approximation so that
the average of the two slightly different coupling constants just 
corresponds to $J^{''}$.
The magnitude of each coupling constant can not be
determined separately from the present analysis.

\section*{Acknowledgments}

The authors would like to thank H. Kageyama, K. Totsuka, H. Nojiri
M. Takigawa, and K. Kodama for many helpful discussions.
The calculations were performed on the HITACHI SR2201 massively
parallel computer of the University of Tokyo and of the
Center for Promotion of Computational Science and Engineering
of Japan Atomic Energy Research Institute.

\end{document}